\documentclass[twocolumn,preprintnumbers,floatfix,prb,superscriptaddress]{revtex4}
\usepackage{graphicx}
\usepackage{graphicx}
\usepackage{dcolumn} 
\usepackage{bm}
\usepackage{epsfig}
\begin{document} 

\title{Pressure-induced metal-insulator and spin-state transition in low-valence layered nickelates}

\author{Victor Pardo}
 \email{vpardo@ucdavis.edu}
\affiliation{Department of Physics,
  University of California, Davis, CA 95616
}

\author{Warren E. Pickett}
 \email{wepickett@ucdavis.edu}
\affiliation{Department of Physics,
  University of California, Davis, CA 95616
}

\date{\today}

\begin{abstract}

Ab initio calculations predict a metal-insulator transition at zero temperature to occur in La$_4$Ni$_3$O$_8$ at moderate pressures as a result of a pressure-induced spin-state transition. The spin-state transition that is seen at 105 K at ambient pressure from a low-temperature high-spin state to a high-temperature low-spin state has been observed to be shifted to lower temperatures as pressure is applied. From our calculations we find that a smaller unit cell volume favors the metallic low-spin state, which becomes more stable at 5 GPa. Similar physics should take place in the related compound La$_3$Ni$_2$O$_6$, but on a different energy scale, which may account for why the transition has not been observed in this material.

\end{abstract}

\maketitle

\section{Introduction}

Spin-state transitions are observed in a variety of systems as a result of the competition between Hund's rule coupling (J$_H$) and crystal field strength ($\Delta_{cf}$). Typical examples are LaCoO$_3$, where Co$^{3+}$:d$^6$ cations in an octahedral environment can occur in a non-magnetic low-spin (LS) state and various excited magnetic spin-states (both intermediate [IS] and high spin [HS] have been predicted\cite{lacoo3}) due to J$_H$ of the cation being the same order of magnitude as $\Delta_{cf}$. Various Fe compounds also show spin-state transitions, and this is a common feature in organometallics literature,\cite{organics_sst} where metal cations can be tuned to be in different spin states. It is common that the LS state is more stable at lower temperatures, but in some systems, the opposite occurs.\cite{invar_alloys} Important competition between $\Delta_{cf}$ and $J_H$ at high pressure has also been found in calculation of the Mott transition in MnO, where the insulator-metal transition, moment collapse, and volume collapse are found (experimentally and theoretically) just above 100 GPa.  The pressure at which the transition occurs is sensitive to competition between these two energy scales, and not to the correlation strength to bandwidth ratio.\cite{kunesMnO}

La$_4$Ni$_3$O$_8$ is a low-valence nickelate that has recently drawn some attention\cite{lanio_curro_1} because of the similarities in its crystallographic and electronic structure with superconducting cuprates. 
It crystallizes in a layered structure\cite{la4ni3o8_struct} formed by three NiO$_2$ neighboring planes, with these trilayers separated by fluorite La/O$_2$/La layers, providing a highly two-dimensional electronic structure. This compound is obtained from La$_4$Ni$_3$O$_{10}$ (Ref. \onlinecite{la4ni3o10_greenblatt_jssc}) by eliminating the apical oxygens neighboring Ni cations, thus the environment changes from octahedral NiO$_6$ in metallic La$_4$Ni$_3$O$_{10}$ to square planar NiO$_4$ in insulating La$_4$Ni$_3$O$_{8}$. Two different Ni sites exist in the structure, depending on whether Ni is in the inner layer or the outer layers of the NiO$_2$ trilayers. The compound undergoes a phase transition around 105 K that has been described via $^{139}$La nuclear magnetic resonance measurements as a transition to an unconventional low-temperature antiferromagnetic (AF) phase.\cite{lanio_curro_2} Recent structural studies\cite{lokshin_mm} suggest the transition is due to a spin-state transition on the Ni cations.

\begin{figure}[ht]
\begin{center}
\includegraphics[width=\columnwidth,draft=false]{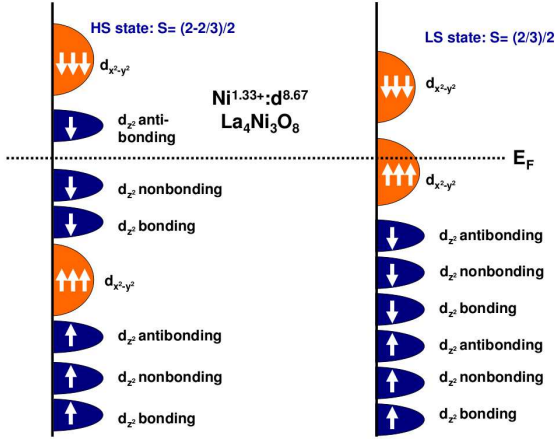}
\caption{Two possible spin states in the trilayer Ni compound La$_4$Ni$_3$O$_8$ are represented depending on the value of the splitting between the d$_{z^2}$ and d$_{x^2-y^2}$ orbitals and the Hund's rule coupling strength. On the left side, we see the high-spin state (larger Hund's rule coupling) and the low-spin state on the right (larger crystal field splitting). Three Ni atoms are represented to account for the strong bonding along the c-axis and the formation of molecular orbitals with d$_{z^2}$ parentage. The Fermi level is represented showing the insulating HS vs. metallic LS solution.}\label{es}
\end{center}
\end{figure}

Two possible spin states can occur in La$_4$Ni$_3$O$_8$. The (on average) Ni$^{1.33+}$:d$^{8.67}$ cations sit in a square planar environment that leads to a large splitting between the d$_{x^2-y^2}$ and d$_{z^2}$ bands caused by the absence of apical oxygens in this low-valence nickelate. $\Delta_{cf}$ within the e$_g$ doublet can be then comparable to J$_H$; if the former is larger, a LS state develops and if the latter is larger, a HS state would be more stable. This distinction is depicted in Fig. \ref{es}. This simple model of the electronic structure of the compound suggests that the HS state would have a larger in-plane lattice parameter due to the slightly larger occupation of the d$_{x^2-y^2}$ orbital and also a smaller Ni-Ni inter-plane distance due to the de-occupation of the d$_{z^2}$ antibonding orbital, as has been recently measured.\cite{lokshin_mm} Moreover, the HS state leads to an in-plane AF coupling being more stable,\cite{la4ni3o8_vpardo} consistent with the observations of its magnetic properties.

As can be seen in Fig. \ref{es}, the HS and LS states would lead to drastically different properties. The HS ion leads to an insulating solution\cite{la4ni3o8_vpardo} caused by the formation of d$_{z^2}$ molecular orbital states. These are bonding-antibonding split around the Fermi level, due to their strong $\sigma$-bond along the c-axis. However, the LS state would have a 2/3-filled d$_{x^2-y^2}$ band around the Fermi level, leading to a metallic result. Experimentally,\cite{lanio_curro_1} a reduction in resistivity above the phase transition at 105 K has been observed, which is consistent with at least an admixture of some Ni atoms in a LS state occurring at high temperatures. If all Ni atoms were in a LS state, the system would become metallic.

\begin{figure*}[ht]
\begin{center}
\includegraphics[width=\columnwidth,draft=false]{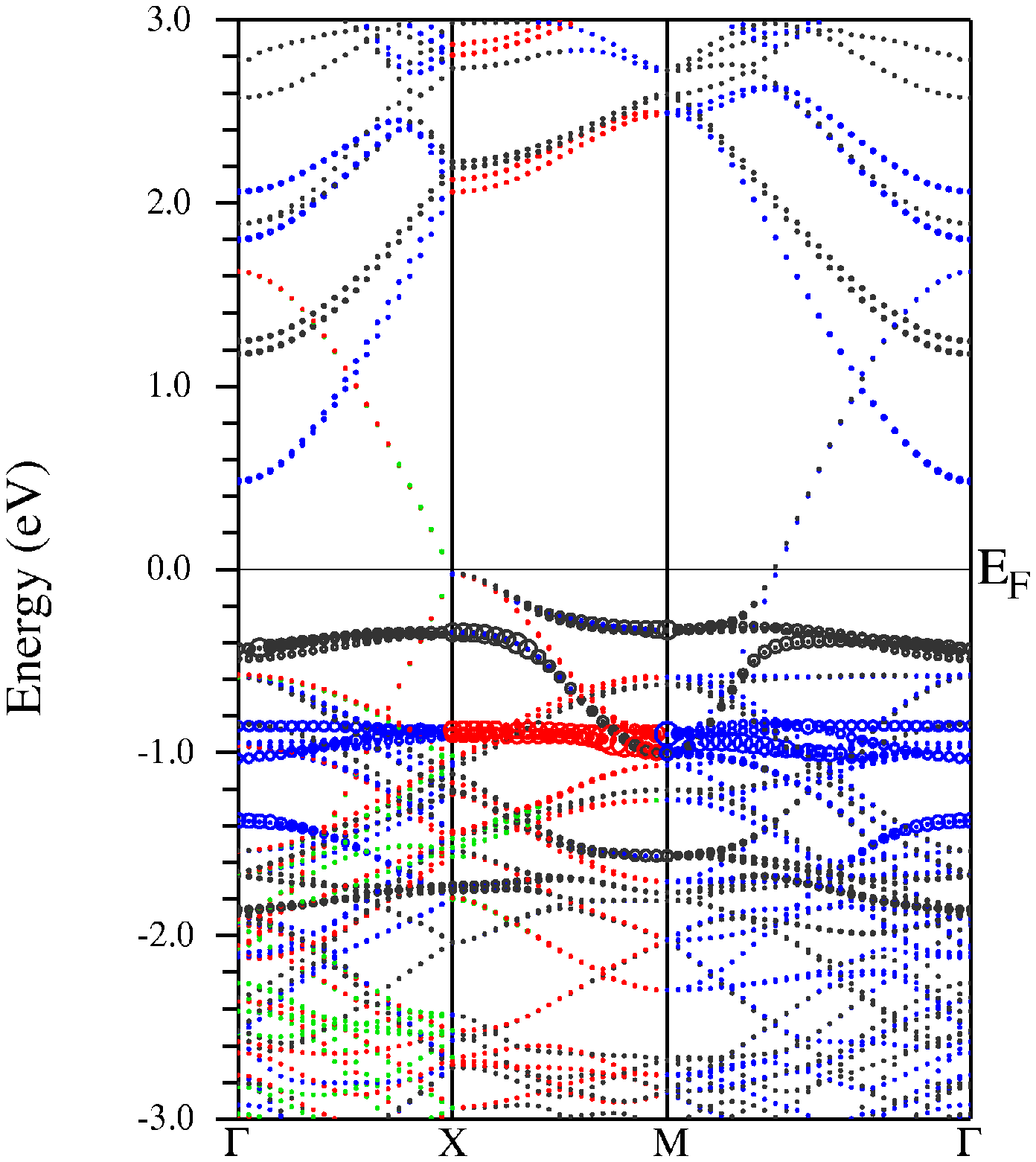}
\includegraphics[width=\columnwidth,draft=false]{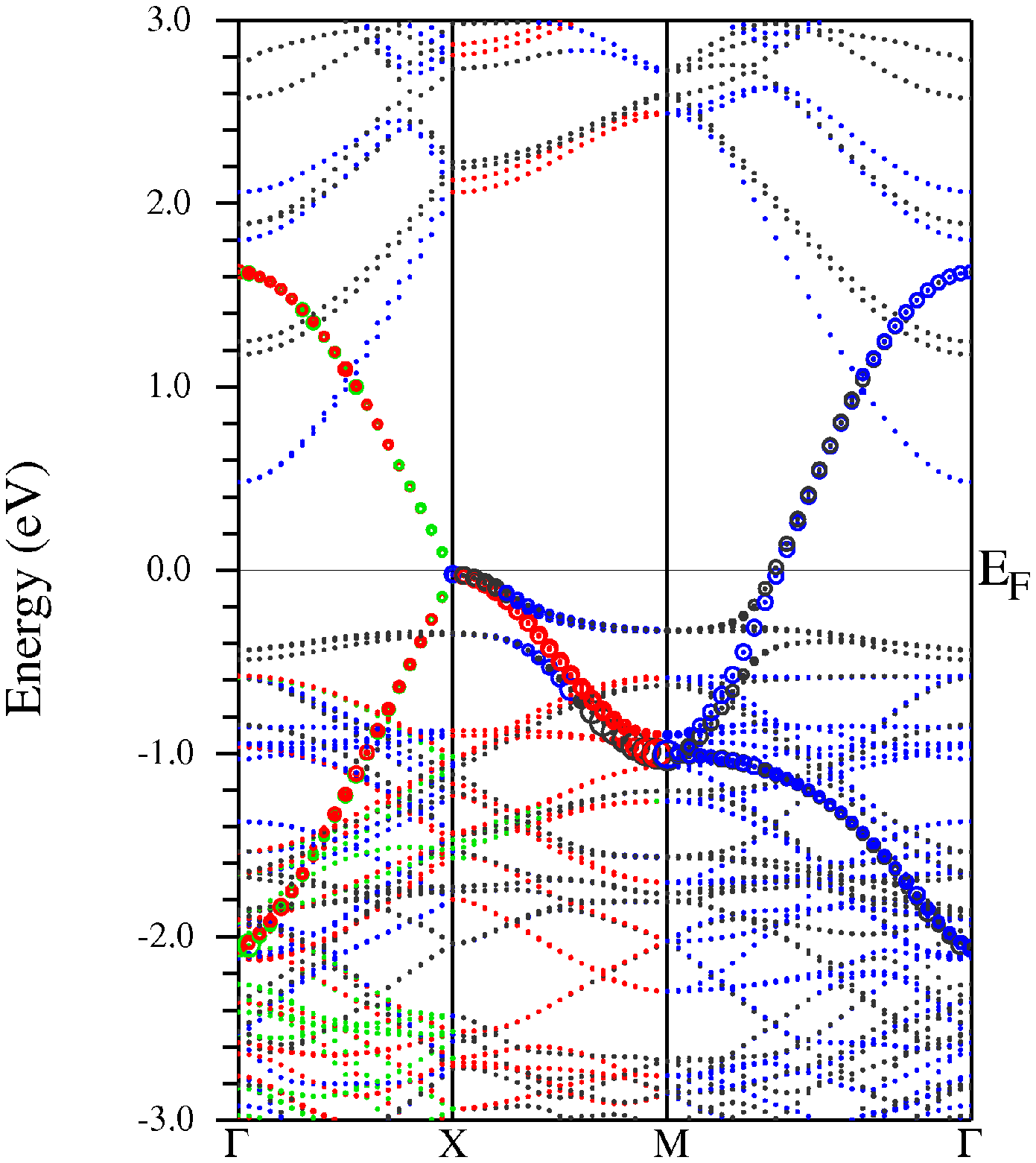}
\caption{Band structures of the LS state are shown, calculated with LDA+U (around the mean field) for U= 6.8 eV. On the left (right) panel, the d$_{z^2}$ (d$_{x^2-y^2}$) bands are highlighted. 2/3-filled d$_{x^2-y^2}$ crosses the Fermi level leading to a metallic solution.}\label{bs_ls}
\end{center}
\end{figure*}

La$_4$Ni$_3$O$_8$ and La$_3$Ni$_2$O$_6$ (similar, but containing NiO$_2$ bilayers instead of trilayers) are compounds close to a metal-insulator transition, and can be driven metallic by oxygen doping maintaining the same structure.\cite{lanio_jpsj} This transition has been recently analyzed for the La$_3$Ni$_2$O$_{7-\delta}$ series.\cite{la3ni2o7_vpardo} For the case of La$_4$Ni$_3$O$_8$, it has been shown that even though the resistivity is quite high and insulating-like, the NMR data has a metallic-like contribution coming from spin-spin scattering with quasiparticles at the Fermi surface (a constant term in the 1/T$_1$T vs. T data\cite{lanio_curro_2} fit for temperatures above the transition at 105 K). In this paper, we explore the evolution of both spin states and their relative stabilizations, with respect to the different computational and physical parameters (values of U, LDA+U scheme, volume variations) suggesting that a spin-state transition can also be attained, with the corresponding reduction in resistivity, at moderate pressures.

\section{Computational Details}

Our electronic structure calculations were  performed within density functional
theory\cite{dft,dft_2} using the all-electron, full potential code {\sc wien2k}\cite{wien}
based on the augmented plane wave plus local orbital (APW+lo) basis set.\cite{sjo}
The generalized gradient approximation\cite{gga} (GGA) was used for the structure optimizations at each volume (that include both optimizations of the $c/a$ ratio and the atomic positions) to estimate the equation of state of the system. Pressure values were obtained utilizing the Murnaghan\cite{murnaghan} and also the Birch-Murnaghan\cite{birch_murnaghan} equations of state.
To deal with  strong correlation effects we apply the LDA+U scheme \cite{sic1,sic2} including an on-site repulsion U and Hund's coupling J 
for the Ni $3d$ states. Results presented below compare two possible LDA+U schemes: the so-called ``fully localized limit"\cite{fll} (FLL) and the ``around the mean field" (AMF) scheme.\cite{amf} A description of the results obtained at different values of U (in a reasonably broad range 4.5-8.5 eV for the Ni cations) is given in the main text below. The value chosen for the on-site Hund's rule strength is J = 0.68 eV.

\section{Results}

\subsection{La$_4$Ni$_3$O$_8$}

The electronic structure of the HS state has been described extensively elsewhere.\cite{la4ni3o8_vpardo,la3ni2o7_vpardo} Here we present the band structures of the LS state. This state, as has been calculated before,\cite{lanio_curro_1} has an in-plane ferromagnetic (FM) coupling due to the less than half-filled d$_{x^2-y^2}$ in-plane orbital, and the coupling out of the plane is small and AF. In Fig. \ref{bs_ls} we present the band structure for one of the spin channels, with the ``fat-bands" highlighting the e$_g$ orbitals of the outer Ni atoms. The electronic structure in this phase was sketched in the right panel of Fig. \ref{es}, where the d$_{z^2}$ bands in both spin channels are shown to be fully occupied. The left panel of Fig. \ref{bs_ls} shows the minority-spin Ni d$_{z^2}$ bands highlighted. These are split into the bonding, non-bonding and antibonding bands. As described in detail in Ref. \onlinecite{la4ni3o8_vpardo} the inner Ni atoms do not contribute to the non-bonding d$_{z^2}$ molecular state (in Fig. \ref{bs_ls} we have highlighted only the outer Ni for simplicity). We can observe that the partly filled d$_{x^2-y^2}$ bands cross the Fermi level leading to a metallic state (these are almost degenerate for both the outer and inner Ni atoms), their bandwidth being more than 3.5 eV. In previous works where the band structure of the HS state is presented,\cite{la4ni3o8_vpardo,la3ni2o7_vpardo} the d$_{x^2-y^2}$ bands are narrower, their bandwidth reduced by the AF in-plane coupling which is not stable in the case of a LS state due to the different filling of the d$_{x^2-y^2}$ band. Dispersions along k$_z$ are negligible in this highly two-dimensional compound. Hence, only the two-dimensional tetragonal Brillouin zone is analyzed.

\begin{figure}[ht]
\begin{center}
\includegraphics[width=\columnwidth,draft=false]{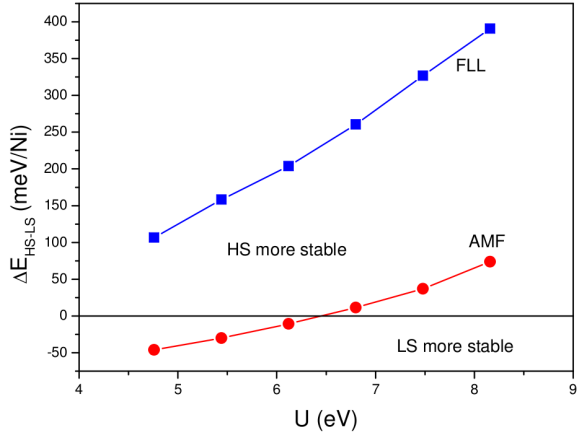}
\caption{Evolution of the total energy difference between the HS and the LS states with U for the two types of LDA+U schemes we have used: FLL (fully localized limit) and AMF (around the mean field). Positive (negative) energy differences indicate the HS (LS) state is more stable. A crossover is found for the AMF scheme at around U= 6.5 eV. }\label{energies}
\end{center}
\end{figure}

We have studied the energetics of the two possible spin states. Figure \ref{energies} shows our analysis of the total energies we have computed. We present the results of the total energy difference between the two spin states as a function of U (in a reasonable range of values from 4.5 to 8.5 eV) for the two LDA+U schemes used (FLL and AMF). FLL predicts the HS state to be more stable for all values of U. Its stability increases as U is increased. The AMF LDA+U scheme, however, tends to favor the stabilization of LS states.\cite{sic2} This is observed in Fig. \ref{energies}, with the AMF method it is possible to obtain similar stabilization energies for the two spin states, whereas in the FLL scheme, an unrealistically small value of U would be necessary. As we can see from our calculations, it is difficult to describe the energetics of the different spin states in La$_4$Ni$_3$O$_8$ purely ab initio, we need to make additional assumptions. From experiments, a spin-state transition is observed in this compound at 105 K, to a low-temperature stabilization of the HS state.\cite{lokshin_mm} Assuming that at temperatures above the transition the LS state can be thermally accessed, we could infer that the order of magnitude of the energy difference between both spin states is about 105 K ($\sim$ 10 meV/Ni). Accepting this, one can use the AMF scheme with a U $\sim$ 7 eV as a correct description of the energetics of the spin state transition for this particular compound. This is the value and scheme utilized for the band structures presented in Fig. \ref{bs_ls}.

It is of special interest to study which spin state becomes favored when pressure is applied to the system. For doing this, we utilize the AMF scheme with U= 6.8 eV, a value that reproduces the energetics of the transition correctly, as we have just described above. It is difficult to estimate a priori which spin state will be favored at high pressure. The energies coming into play will be the relative position of the two e$_g$ states ($\Delta_{cf}$) and the bonding-antibonding splitting in the d$_{z^2}$ bands. Applying pressure reduces the Ni-Ni interplanar distance, this enhances the bonding-antibonding splitting (further promoting a HS state), but also lowers the relative energy of the d$_{z^2}$ band (a stronger metal-metal bond along the c-axis produces a larger d$_{z^2}$ occupation to screen the repulsion). Within the plane, reduction of the Ni-Ni distance destabilizes the d$_{x^2-y^2}$ orbital favoring the LS state. All in all, several energies need to be taken into account and thus it is not easy to anticipate what pressure effects will do to this system before doing the calculations. 

\begin{figure}[ht]
\begin{center}
\includegraphics[width=\columnwidth,draft=false]{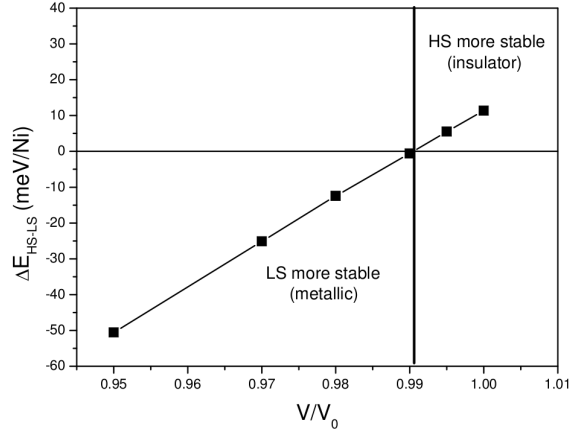}
\caption{Evolution of the total energy difference between the HS and the LS states as a function of volume. LS becomes more stable at lower volumes. Calculations were carried out with the AMF LDA+U scheme for U= 6.8 eV (see text).}\label{pressure}
\end{center}
\end{figure}

Results from the actual total energy calculations are summarized in Fig. \ref{pressure}. These show a reduction in volume favors the LS state. The calculations were carried out using the same atomic positions as observed experimentally\cite{la4ni3o8_struct} and the same c/a value for all volumes considered. We observe a very rapid variation in the total energy difference, suggesting a transition from a HS to a LS state occurs at relatively small pressures. As we have discussed above, the LS state would lead to a metallic solution, with a 2/3-filled d$_{x^2-y^2}$ band crossing the Fermi level. Thus a metal-insulator transition is predicted from our calculations (or at least the accessibility of LS states, that would eventually lead to metallic behavior) at a modest volume reduction of only about 1\% with respect to the experimental one.

To calculate the pressure at which we predict the transition to take place, we have carried out a series of GGA calculations for the structural optimization at each volume. GGA is known to give good estimates of the lattice constants, and this is generally not improved by the LDA+U method. Our GGA calculations predict a unit cell volume 1\% larger than the experimental data (the theoretical lattice parameters are $a$= 3.95 \AA, c= 26.42 \AA, in good agreement with the experimental\cite{lanio_curro_1} $a$= 3.96 \AA, c= 26.04 \AA). We have fit our total energies calculated for each volume to both a Murnaghan\cite{murnaghan} and a Birch-Murnaghan\cite{birch_murnaghan} equation, to obtain the pressure vs. volume relationship. The crossover between HS and LS, with the corresponding pressure-induced metal-insulator transition takes place at about 5 GPa, the bulk modulus of the compound being 180 GPa.



\subsection{Comparison with La$_3$Ni$_2$O$_6$}

\begin{figure*}[ht]
\begin{center}
\includegraphics[width=\columnwidth,draft=false]{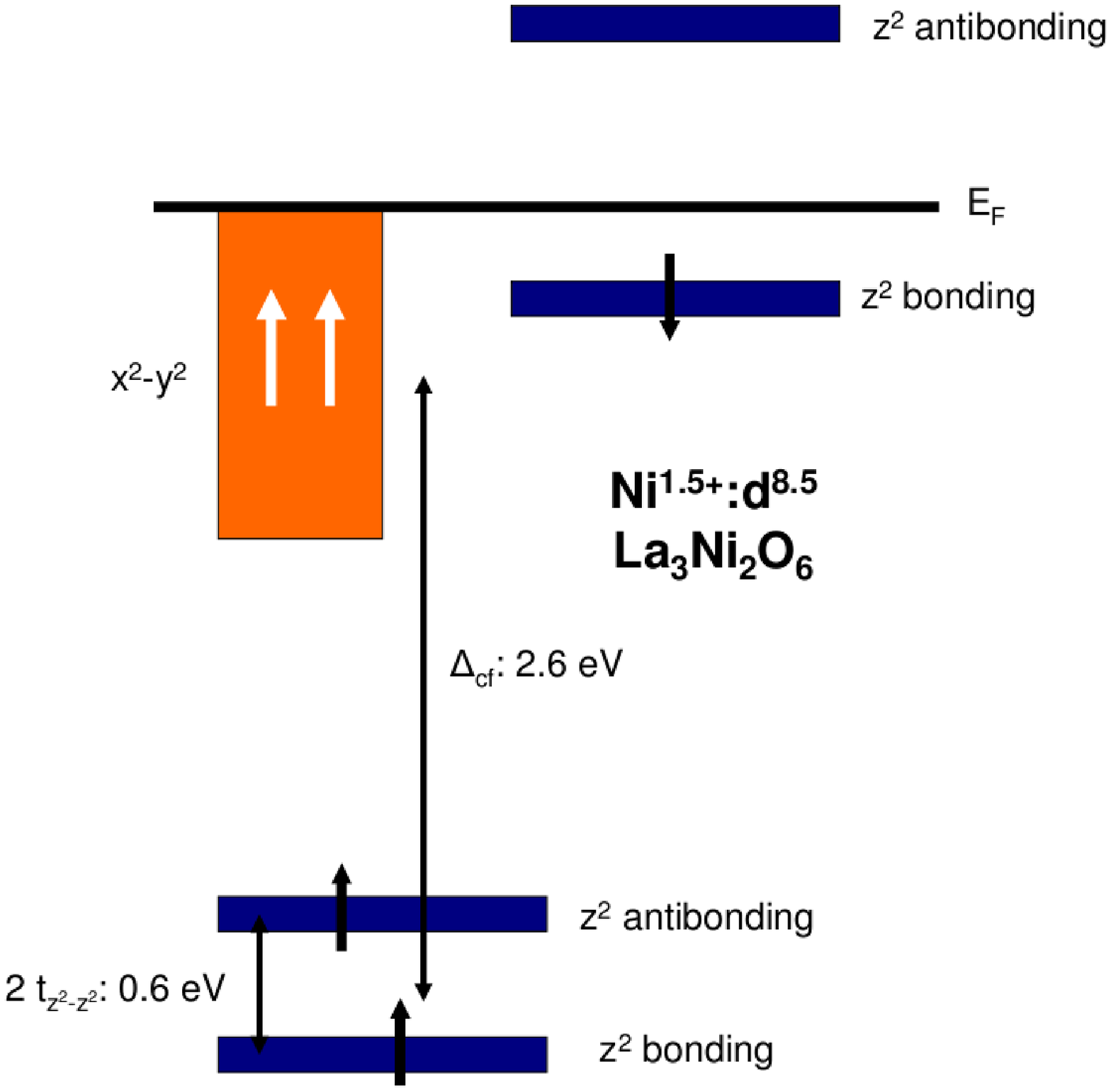}
\includegraphics[width=\columnwidth,draft=false]{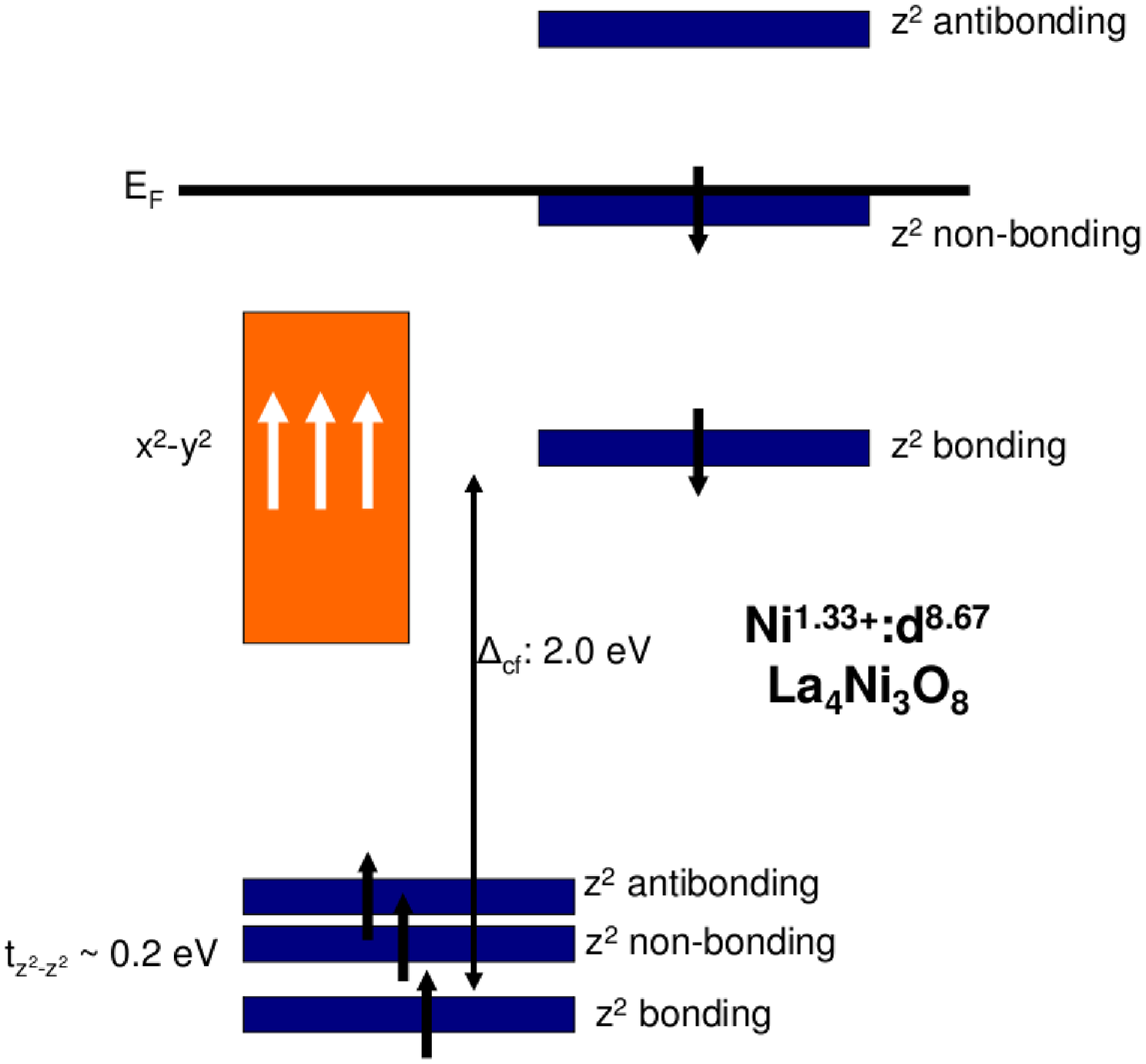}
\caption{Comparison of the band structure parameters obtained for La$_4$Ni$_3$O$_8$ and La$_3$Ni$_2$O$_6$ in Refs. \onlinecite{la4ni3o8_vpardo} and \onlinecite{la3ni2o7_vpardo} from LDA+U calculations. Observe that a smaller crystal-field splitting and a larger bonding-antibonding splitting occur for La$_3$Ni$_2$O$_6$ compared to La$_4$Ni$_3$O$_8$. Three (two) Ni atoms are represented for the case of La$_4$Ni$_3$O$_8$ (La$_3$Ni$_2$O$_6$) to account for the bonding-antibonding splitting of the d$_{z^2}$ bands.}\label{cf}
\end{center}
\end{figure*}

\begin{figure}[ht]
\begin{center}
\includegraphics[width=\columnwidth,draft=false]{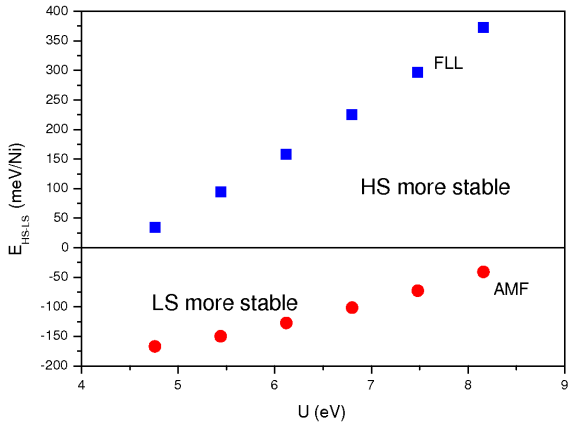}
\caption{Evolution of the total energy difference between the HS and the LS states with U for the two types of LDA+U schemes we have used: FLL (fully localized limit) and AMF (around the mean field) for La$_3$Ni$_2$O$_6$. Positive (negative) energy differences indicate the HS (LS) state is more stable. Comparing with the results for La$_4$Ni$_3$O$_8$, the LS state is closer in energy, being always favored in the AMF scheme.}\label{326}
\end{center}
\end{figure}

In principle, the physics of the spin-state transition is the same in this compound and in La$_3$Ni$_2$O$_6$. However, the latter does not show experimentally a spin-state transition,\cite{la3ni2o6_prl} or this happens in a smoother way with no signatures of a transition in susceptibility, resistivity and heat capacity. Our picture would predict the same metal-insulator transition to occur on a different pressure/temperature scale, but the lack of experimental input about the energetics of the two possible spin states prevents us from being able to estimate the pressure required for the transition. For the same reason that it is difficult to predict what direction pressure will drive the relative stabilization of a HS vs. LS state, it is difficult to compare the situation in La$_3$Ni$_2$O$_6$ with La$_4$Ni$_3$O$_8$. If we compare the band structures presented with the same value of U and the same LDA+U scheme (FLL) in Refs. \onlinecite{la4ni3o8_vpardo} and \onlinecite{la3ni2o7_vpardo} for the HS state, one can extract from them the band structure parameters ($t_{d_{z^2}-d_{z^2}}$ being the hopping parameter for the $\sigma$-bond between Ni cations along the c-axis and $\Delta_{cf}$ the intra-e$_g$ crystal field splitting), that are sketched in Fig. \ref{cf}. $t_{d_{z^2}-d_{z^2}}$ $\sim$ 0.3 eV, $\Delta_{cf}$ $\sim$ 2.6 eV for La$_3$Ni$_2$O$_6$, and $t_{d_{z^2}-d_{z^2}}$ $\sim$ 0.2 eV, $\Delta_{cf}$ $\sim$ 2.0 eV for La$_4$Ni$_3$O$_8$, which is consistent with their similar structural parameters (3.96 \AA\ for the Ni-Ni in-plane distance for both materials and a slightly larger out-of-plane Ni-Ni distance for La$_4$Ni$_3$O$_8$: 3.25 vs. 3.19 \AA). The larger crystal field splitting in La$_3$Ni$_2$O$_6$ would favor the LS state with respect to the case of La$_4$Ni$_3$O$_8$, however the larger bonding-antibonding splitting in La$_3$Ni$_2$O$_6$ goes in the opposite direction favoring the HS state also. Thus while similar physical properties are highly expected in both compounds, the exact energetics of a possible transition in La$_3$Ni$_2$O$_6$ are difficult to predict purely ab initio. One big difference is the existence of a non-bonding d$_{z^2}$ band in La$_4$Ni$_3$O$_8$ just 0.3 eV above the top of the occupied d$_{x^2-y^2}$ band (see right panel of Fig. \ref{cf}), whereas in La$_3$Ni$_2$O$_6$ the antibonding d$_{z^2}$ band is significantly higher in energy, but the top of the valence band has d$_{x^2-y^2}$ character in that case (see left panel of Fig. \ref{cf}).

To obtain a more qualitative understanding of a possible spin-state transition in La$_3$Ni$_2$O$_6$, we have computed the total energy difference between the HS and LS state for the same two LDA+U schemes we used for La$_4$Ni$_3$O$_8$ with the same range of U values and also in the experimental structure.\cite{la3ni2o6_struct} In principle, the slightly smaller d-occupation (nominally d$^{8.5}$ vs. d$^{8.67}$ in La$_4$Ni$_3$O$_8$) suggests a very similar but slightly smaller value of U should be reasonable for this compound. The results are summarized in Fig. \ref{326}, where by comparison with Fig. \ref{energies} for La$_4$Ni$_3$O$_8$, it can be seen that the LS state should be more easily accessible for La$_3$Ni$_2$O$_6$, suggesting that the larger crystal field splitting plays the biggest role in determining the relative stabilization of the different spin states. Using the AMF scheme, the LS state is always more favored, and the stabilization energy of the HS state in FLL is smaller than in the case of La$_4$Ni$_3$O$_8$. However, no experimental indications to date suggest a spin-state transition takes place in La$_3$Ni$_2$O$_6$: neither specific heat\cite{lanio_curro_1} nor resistivity\cite{la3ni2o6_prl} show any anomalies at a particular temperature. Still, more experiments are required to clarify the different properties of La$_3$Ni$_2$O$_6$ compared to La$_4$Ni$_3$O$_8$.


\section{Summary}

Our ab initio calculations for the compound La$_4$Ni$_3$O$_8$ describe the energetics of the spin-state transition predicted for the system. A low-temperature HS state has been observed experimentally and is predicted to be the ground state by our calculations at room pressure. These suggest the metallic LS state would be stable at moderate pressures ($\sim$ 5 GPa). Our electronic structure model for the system predicts a pressure-induced metal-insulator transition would occur together with the spin-state transition. Also, our description of the electronic structure can account for the different spin-state properties in La$_3$Ni$_2$O$_6$, the changes in lattice parameters in La$_4$Ni$_3$O$_8$ at low temperature, and gives indications of their proximity to a metallic state (for which some experimental evidence already exists for La$_4$Ni$_3$O$_8$).

\acknowledgments
The investigations presented in this paper grew out of questions posed by D.I. Khomskii about which spin state becomes favored at high pressure. The authors acknowledge the exchange of results in the same family of compounds with K. A. Lokshin, J. W. Freeland, C. H. Yee, and G. Kotliar, most of them prior to their publication. This project was supported by Department of Energy grant DE-FG02-04ER46111.


\begin{thebibliography}{25}
\expandafter\ifx\csname natexlab\endcsname\relax\def\natexlab#1{#1}\fi
\expandafter\ifx\csname bibnamefont\endcsname\relax
  \def\bibnamefont#1{#1}\fi
\expandafter\ifx\csname bibfnamefont\endcsname\relax
  \def\bibfnamefont#1{#1}\fi
\expandafter\ifx\csname citenamefont\endcsname\relax
  \def\citenamefont#1{#1}\fi
\expandafter\ifx\csname url\endcsname\relax
  \def\url#1{\texttt{#1}}\fi
\expandafter\ifx\csname urlprefix\endcsname\relax\def\urlprefix{URL }\fi
\providecommand{\bibinfo}[2]{#2}
\providecommand{\eprint}[2][]{\url{#2}}

\bibitem[{\citenamefont{Se{\~n}ar{\'i}s-Rodr{\'i}guez and
  Goodenough}(1995)}]{lacoo3}
\bibinfo{author}{\bibfnamefont{M.~A.}
  \bibnamefont{Se{\~n}ar{\'i}s-Rodr{\'i}guez}} \bibnamefont{and}
  \bibinfo{author}{\bibfnamefont{J.~B.} \bibnamefont{Goodenough}},
  \bibinfo{journal}{J. Solid State Chem.} \textbf{\bibinfo{volume}{116}},
  \bibinfo{pages}{224} (\bibinfo{year}{1995}).

\bibitem[{\citenamefont{G{\"u}tlich et~al.}(2000)\citenamefont{G{\"u}tlich,
  Garc{\'i}a, and Goodwin}}]{organics_sst}
\bibinfo{author}{\bibfnamefont{P.}~\bibnamefont{G{\"u}tlich}},
  \bibinfo{author}{\bibfnamefont{Y.}~\bibnamefont{Garc{\'i}a}},
  \bibnamefont{and} \bibinfo{author}{\bibfnamefont{H.~A.}
  \bibnamefont{Goodwin}}, \bibinfo{journal}{Chem. Soc. Rev.}
  \textbf{\bibinfo{volume}{29}}, \bibinfo{pages}{419} (\bibinfo{year}{2000}).

\bibitem[{\citenamefont{Moruzzi}(1990)}]{invar_alloys}
\bibinfo{author}{\bibfnamefont{V.~L.} \bibnamefont{Moruzzi}},
  \bibinfo{journal}{Phys. Rev. B} \textbf{\bibinfo{volume}{41}},
  \bibinfo{pages}{6939} (\bibinfo{year}{1990}).

\bibitem[{\citenamefont{Kunes et~al.}(2008)\citenamefont{Kunes, Lukoyanov,
  Anisimov, Scalettar, and Pickett}}]{kunesMnO}
\bibinfo{author}{\bibfnamefont{J.}~\bibnamefont{Kune$\check{s}$}},
  \bibinfo{author}{\bibfnamefont{A.~V.} \bibnamefont{Lukoyanov}},
  \bibinfo{author}{\bibfnamefont{V.~I.} \bibnamefont{Anisimov}},
  \bibinfo{author}{\bibfnamefont{R.~T.} \bibnamefont{Scalettar}},
  \bibnamefont{and} \bibinfo{author}{\bibfnamefont{W.~E.}
  \bibnamefont{Pickett}}, \bibinfo{journal}{Nat. Mater.}
  \textbf{\bibinfo{volume}{7}}, \bibinfo{pages}{198} (\bibinfo{year}{2008}).

\bibitem[{\citenamefont{Poltavets et~al.}(2010)\citenamefont{Poltavets,
  Loshkin, Nevidomskyy, Croft, Tyson, Hadermann, Tendeloo, Egami, Kotliar,
  ApRoberts-Warren et~al.}}]{lanio_curro_1}
\bibinfo{author}{\bibfnamefont{V.~V.} \bibnamefont{Poltavets}},
  \bibinfo{author}{\bibfnamefont{K.~A.} \bibnamefont{Loshkin}},
  \bibinfo{author}{\bibfnamefont{A.~H.} \bibnamefont{Nevidomskyy}},
  \bibinfo{author}{\bibfnamefont{M.}~\bibnamefont{Croft}},
  \bibinfo{author}{\bibfnamefont{T.~A.} \bibnamefont{Tyson}},
  \bibinfo{author}{\bibfnamefont{J.}~\bibnamefont{Hadermann}},
  \bibinfo{author}{\bibfnamefont{G.~V.} \bibnamefont{Tendeloo}},
  \bibinfo{author}{\bibfnamefont{T.}~\bibnamefont{Egami}},
  \bibinfo{author}{\bibfnamefont{G.}~\bibnamefont{Kotliar}},
  \bibinfo{author}{\bibfnamefont{N.}~\bibnamefont{ApRoberts-Warren}},
  \bibnamefont{et~al.}, \bibinfo{journal}{Phys. Rev. Lett.}
  \textbf{\bibinfo{volume}{104}}, \bibinfo{pages}{206403}
  (\bibinfo{year}{2010}).

\bibitem[{\citenamefont{Poltavets
  et~al.}(2006{\natexlab{a}})\citenamefont{Poltavets, Lokshin, Egami, and
  Greenblatt}}]{la4ni3o8_struct}
\bibinfo{author}{\bibfnamefont{V.~V.} \bibnamefont{Poltavets}},
  \bibinfo{author}{\bibfnamefont{K.~A.} \bibnamefont{Lokshin}},
  \bibinfo{author}{\bibfnamefont{T.}~\bibnamefont{Egami}}, \bibnamefont{and}
  \bibinfo{author}{\bibfnamefont{M.}~\bibnamefont{Greenblatt}},
  \bibinfo{journal}{Mater. Res. Bull.} \textbf{\bibinfo{volume}{41}},
  \bibinfo{pages}{955} (\bibinfo{year}{2006}{\natexlab{a}}).

\bibitem[{\citenamefont{Zhang and
  Greenblatt}(1995)}]{la4ni3o10_greenblatt_jssc}
\bibinfo{author}{\bibfnamefont{Z.}~\bibnamefont{Zhang}} \bibnamefont{and}
  \bibinfo{author}{\bibfnamefont{M.}~\bibnamefont{Greenblatt}},
  \bibinfo{journal}{J. Solid State Chem.} \textbf{\bibinfo{volume}{117}},
  \bibinfo{pages}{236} (\bibinfo{year}{1995}).

\bibitem[{\citenamefont{ApRoberts-Warren
  et~al.}(2011)\citenamefont{ApRoberts-Warren, Dioguardi, Poltavets,
  Greenblatt, Klavins, and Curro}}]{lanio_curro_2}
\bibinfo{author}{\bibfnamefont{N.}~\bibnamefont{ApRoberts-Warren}},
  \bibinfo{author}{\bibfnamefont{A.~P.} \bibnamefont{Dioguardi}},
  \bibinfo{author}{\bibfnamefont{V.~V.} \bibnamefont{Poltavets}},
  \bibinfo{author}{\bibfnamefont{M.}~\bibnamefont{Greenblatt}},
  \bibinfo{author}{\bibfnamefont{P.}~\bibnamefont{Klavins}}, \bibnamefont{and}
  \bibinfo{author}{\bibfnamefont{N.~J.} \bibnamefont{Curro}},
  \bibinfo{journal}{Phys. Rev. B} \textbf{\bibinfo{volume}{83}},
  \bibinfo{pages}{014402} (\bibinfo{year}{2011}).

\bibitem[{\citenamefont{Lokshin and Egami}(2011)}]{lokshin_mm}
\bibinfo{author}{\bibfnamefont{K.}~\bibnamefont{Lokshin}} \bibnamefont{and}
  \bibinfo{author}{\bibfnamefont{T.}~\bibnamefont{Egami}}
  (\bibinfo{year}{2011}),
  \urlprefix\url{http://meetings.aps.org/Meeting/MAR11/Event/141968}.

\bibitem[{\citenamefont{Pardo and Pickett}(2010)}]{la4ni3o8_vpardo}
\bibinfo{author}{\bibfnamefont{V.}~\bibnamefont{Pardo}} \bibnamefont{and}
  \bibinfo{author}{\bibfnamefont{W.~E.} \bibnamefont{Pickett}},
  \bibinfo{journal}{Phys. Rev. Lett.} \textbf{\bibinfo{volume}{105}},
  \bibinfo{pages}{266402} (\bibinfo{year}{2010}).

\bibitem[{\citenamefont{Kobayashi et~al.}(1996)\citenamefont{Kobayashi,
  Taniguchi, Kasai, Sato, Nishioka, and Kontani}}]{lanio_jpsj}
\bibinfo{author}{\bibfnamefont{Y.}~\bibnamefont{Kobayashi}},
  \bibinfo{author}{\bibfnamefont{S.}~\bibnamefont{Taniguchi}},
  \bibinfo{author}{\bibfnamefont{M.}~\bibnamefont{Kasai}},
  \bibinfo{author}{\bibfnamefont{M.}~\bibnamefont{Sato}},
  \bibinfo{author}{\bibfnamefont{T.}~\bibnamefont{Nishioka}}, \bibnamefont{and}
  \bibinfo{author}{\bibfnamefont{M.}~\bibnamefont{Kontani}},
  \bibinfo{journal}{J. Phys. Soc. Japan} \textbf{\bibinfo{volume}{65}},
  \bibinfo{pages}{3978} (\bibinfo{year}{1996}).

\bibitem[{\citenamefont{Pardo and Pickett}(2011)}]{la3ni2o7_vpardo}
\bibinfo{author}{\bibfnamefont{V.}~\bibnamefont{Pardo}} \bibnamefont{and}
  \bibinfo{author}{\bibfnamefont{W.~E.} \bibnamefont{Pickett}},
  \bibinfo{journal}{Phys. Rev. B} \textbf{\bibinfo{volume}{in press}}
  (\bibinfo{year}{2011}).

\bibitem[{\citenamefont{Hohenberg and Kohn}(1964)}]{dft}
\bibinfo{author}{\bibfnamefont{P.}~\bibnamefont{Hohenberg}} \bibnamefont{and}
  \bibinfo{author}{\bibfnamefont{W.}~\bibnamefont{Kohn}},
  \bibinfo{journal}{Phys. Rev.} \textbf{\bibinfo{volume}{136}},
  \bibinfo{pages}{B864} (\bibinfo{year}{1964}).

\bibitem[{\citenamefont{Jones and Gunnarsson}(1989)}]{dft_2}
\bibinfo{author}{\bibfnamefont{R.~O.} \bibnamefont{Jones}} \bibnamefont{and}
  \bibinfo{author}{\bibfnamefont{O.}~\bibnamefont{Gunnarsson}},
  \bibinfo{journal}{Rev. Mod. Phys.} \textbf{\bibinfo{volume}{61}},
  \bibinfo{pages}{689} (\bibinfo{year}{1989}).

\bibitem[{\citenamefont{Schwarz and Blaha}(2003)}]{wien}
\bibinfo{author}{\bibfnamefont{K.}~\bibnamefont{Schwarz}} \bibnamefont{and}
  \bibinfo{author}{\bibfnamefont{P.}~\bibnamefont{Blaha}},
  \bibinfo{journal}{Comp. Mat. Sci.} \textbf{\bibinfo{volume}{28}},
  \bibinfo{pages}{259} (\bibinfo{year}{2003}).

\bibitem[{\citenamefont{Sj{\"o}stedt et~al.}(2000)\citenamefont{Sj{\"o}stedt,
  N{\"o}rdstrom, and Singh}}]{sjo}
\bibinfo{author}{\bibfnamefont{E.}~\bibnamefont{Sj{\"o}stedt}},
  \bibinfo{author}{\bibfnamefont{L.}~\bibnamefont{N{\"o}rdstrom}},
  \bibnamefont{and} \bibinfo{author}{\bibfnamefont{D.~J.} \bibnamefont{Singh}},
  \bibinfo{journal}{Solid State Commun.} \textbf{\bibinfo{volume}{114}},
  \bibinfo{pages}{15} (\bibinfo{year}{2000}).

\bibitem[{\citenamefont{Perdew et~al.}(1996)\citenamefont{Perdew, Burke, and
  Ernzerhof}}]{gga}
\bibinfo{author}{\bibfnamefont{J.~P.} \bibnamefont{Perdew}},
  \bibinfo{author}{\bibfnamefont{K.}~\bibnamefont{Burke}}, \bibnamefont{and}
  \bibinfo{author}{\bibfnamefont{M.}~\bibnamefont{Ernzerhof}},
  \bibinfo{journal}{Phys.\ Rev. Lett.} \textbf{\bibinfo{volume}{77}},
  \bibinfo{pages}{3865} (\bibinfo{year}{1996}).

\bibitem[{\citenamefont{Murnaghan}(1944)}]{murnaghan}
\bibinfo{author}{\bibfnamefont{F.~D.} \bibnamefont{Murnaghan}},
  \bibinfo{journal}{Proc. Natl. Acad. Sci. USA} \textbf{\bibinfo{volume}{30}},
  \bibinfo{pages}{244} (\bibinfo{year}{1944}).

\bibitem[{\citenamefont{Birch}(1947)}]{birch_murnaghan}
\bibinfo{author}{\bibfnamefont{F.}~\bibnamefont{Birch}},
  \bibinfo{journal}{Phys. Rev.} \textbf{\bibinfo{volume}{71}},
  \bibinfo{pages}{809} (\bibinfo{year}{1947}).

\bibitem[{\citenamefont{Anisimov et~al.}(1991)\citenamefont{Anisimov, Zaanen,
  and Andersen}}]{sic1}
\bibinfo{author}{\bibfnamefont{V.~I.} \bibnamefont{Anisimov}},
  \bibinfo{author}{\bibfnamefont{J.}~\bibnamefont{Zaanen}}, \bibnamefont{and}
  \bibinfo{author}{\bibfnamefont{O.~K.} \bibnamefont{Andersen}},
  \bibinfo{journal}{Phys. Rev. B} \textbf{\bibinfo{volume}{44}},
  \bibinfo{pages}{943} (\bibinfo{year}{1991}).

\bibitem[{\citenamefont{Ylvisaker et~al.}(2009)\citenamefont{Ylvisaker,
  Pickett, and Koepernik}}]{sic2}
\bibinfo{author}{\bibfnamefont{E.~R.} \bibnamefont{Ylvisaker}},
  \bibinfo{author}{\bibfnamefont{W.~E.} \bibnamefont{Pickett}},
  \bibnamefont{and}
  \bibinfo{author}{\bibfnamefont{K.}~\bibnamefont{Koepernik}},
  \bibinfo{journal}{Phys. Rev. B} \textbf{\bibinfo{volume}{79}},
  \bibinfo{pages}{035103} (\bibinfo{year}{2009}).

\bibitem[{\citenamefont{Petukhov et~al.}(2003)\citenamefont{Petukhov, Mazin,
  Chioncel, and Lichtenstein}}]{fll}
\bibinfo{author}{\bibfnamefont{A.}~\bibnamefont{Petukhov}},
  \bibinfo{author}{\bibfnamefont{I.}~\bibnamefont{Mazin}},
  \bibinfo{author}{\bibfnamefont{L.}~\bibnamefont{Chioncel}}, \bibnamefont{and}
  \bibinfo{author}{\bibfnamefont{A.}~\bibnamefont{Lichtenstein}},
  \bibinfo{journal}{Phys.\ Rev. B} \textbf{\bibinfo{volume}{67}},
  \bibinfo{pages}{153106} (\bibinfo{year}{2003}).

\bibitem[{\citenamefont{Czyzyk and Sawatzky}(1994)}]{amf}
\bibinfo{author}{\bibfnamefont{M.~T.} \bibnamefont{Czyzyk}} \bibnamefont{and}
  \bibinfo{author}{\bibfnamefont{G.~A.} \bibnamefont{Sawatzky}},
  \bibinfo{journal}{Phys. Rev. B} \textbf{\bibinfo{volume}{49}},
  \bibinfo{pages}{14211} (\bibinfo{year}{1994}).

\bibitem[{\citenamefont{Poltavets et~al.}(2009)\citenamefont{Poltavets,
  Greenblatt, Fecher, and Felser}}]{la3ni2o6_prl}
\bibinfo{author}{\bibfnamefont{V.~V.} \bibnamefont{Poltavets}},
  \bibinfo{author}{\bibfnamefont{M.}~\bibnamefont{Greenblatt}},
  \bibinfo{author}{\bibfnamefont{G.~H.} \bibnamefont{Fecher}},
  \bibnamefont{and} \bibinfo{author}{\bibfnamefont{C.}~\bibnamefont{Felser}},
  \bibinfo{journal}{Phys. Rev. Lett.} \textbf{\bibinfo{volume}{102}},
  \bibinfo{pages}{046405} (\bibinfo{year}{2009}).

\bibitem[{\citenamefont{Poltavets
  et~al.}(2006{\natexlab{b}})\citenamefont{Poltavets, Loshin, Dikmen, Croft,
  Egami, and Greenblatt}}]{la3ni2o6_struct}
\bibinfo{author}{\bibfnamefont{V.~V.} \bibnamefont{Poltavets}},
  \bibinfo{author}{\bibfnamefont{K.~A.} \bibnamefont{Loshin}},
  \bibinfo{author}{\bibfnamefont{S.}~\bibnamefont{Dikmen}},
  \bibinfo{author}{\bibfnamefont{M.}~\bibnamefont{Croft}},
  \bibinfo{author}{\bibfnamefont{T.}~\bibnamefont{Egami}}, \bibnamefont{and}
  \bibinfo{author}{\bibfnamefont{M.}~\bibnamefont{Greenblatt}},
  \bibinfo{journal}{J. Am. Chem. Soc.} \textbf{\bibinfo{volume}{128}},
  \bibinfo{pages}{9050} (\bibinfo{year}{2006}{\natexlab{b}}).

\end{thebibliography}

\end{document}